# Ultrafast photochromism and bacteriochromism in one dimensional hybrid plasmonic photonic structures

Francesco Scotognella*[a,b], Giuseppe M. Paternò[b], Ilka Kriegel[c], Silvio Bonfadini[b], Liliana Moscardi[a,b], Luigino Criante[b], Stefano Donini[b], Davide Ariodanti[a], Maurizio Zani[a], Emilio Parisini[b], Guglielmo Lanzani[a]

[a]Dipartimento di Fisica, Politecnico di Milano, Piazza Leonardo da Vinci 32, 20133 Milano, Italy;
[b]Center for Nano Science and Technology@PoliMi, Istituto Italiano di Tecnologia, Via Giovanni Pascoli, 70/3, 20133, Milan, Italy; [c]Functional Nanosystems, Istituto Italiano di Tecnologia (IIT), via Morego, 30, 16163 Genova, Italy

## ABSTRACT

Hybrid plasmonic photonic structures combine the plasmonic response with the photonic band gap, holding promise for utilization as optical switches and sensors. Here, we demonstrate the active modulation of the optical response in such structures with two different external stimuli, *e.g.* laser pulses and bacteria.

First, we report the fabrication of a miniaturized (5 x 5 mm) indium tin oxide (ITO) grating employing femtosecond laser micromachining, and we show the possibility to modulate the photonic band gap in the visible via ultrafast photoexcitation in the infrared part of the spectrum. Note that the demonstrated time response in the picosecond range of the spectral modulation have an industrial relevance.

Moreover, we manufacture one-dimensional photonic crystals consisting of a solution-processed dielectric Bragg stack exposing a top-layer of bio-active silver. We assign the bacterial responsivity of the system to polarization charges at the Ag/bacterium interface, giving rise to an overall blue shift of the photonic band gap.

*francesco.scotognella@polimi.it; phone 39 02 2399-6056; fax 39 02 2399-6126; www.fisi.polimi.it/en/people/scotognella

## 1. INTRODUCTION

The physical properties of photonic crystals, after their discovery in 1987 [1,2], are still very interesting and attract the interest of the scientific community [3–7]. The main characteristic of a photonic crystal is the space periodicity of the dielectric permittivity and, thus, of the refractive index, that. results in energy regions in which light propagation is not allowed, the so-called photonic band gaps [8–10]. The dielectric permittivity periodicity can be achieved in one, two and three dimensions. The one-dimensional case, is the simplest and it can be obtained in several ways. For example, a substrate can be dug to obtain a pattern of lines of two different materials [8,11]. Alternatively, one-dimensional photonic crystals can be fabricated by stacking two different types of materials, and these multilayer geometries are usually called Bragg stacks [12–17]. Bragg stacks are very versatile from a materials point of view because they can be fabricated either with organic [18–20] or inorganic materials [21–24], exploiting a wide range of refractive index values. If a material that is employed to fabricate the photonic crystal is a plasmonic, e.g. either a metal or a doped semiconductor nanocrystal [25,26], the resulting hybrid plasmonic photonic crystal combines the photonic band gap with the plasmonic resonance of the suitable material [27–31].

The first example of refractive index periodic modulation, and thus of the photonic band gap control, we discuss is the fabrication of one-dimensional gratings by femtosecond laser machining. The grating has been obtained by alternating lines of ITO and air. The advantage of femtosecond laser machining is the possibility to fabricate large-area gratings, in this study 5x5 millimeters, in a reasonable time and without masks. By pumping the grating with short ultraviolet pulses

(with photon energy above the electronic band gap of ITO) we are increasing the number of carriers in the conduction band of ITO, resulting in a large plasma frequency. The change in the plasma frequency induces a change in the dielectric permittivity and, thus, the refractive index of ITO, leading to a modification of the effective refractive index of the ITO grating and a shift of the grating photonic band gap. The behavior of the ITO grating become very interesting when pumped with near-infrared laser pulses, whose energy is below the electronic band gap but resonant with the ITO plasmonic feature. Such intraband transition can still lead to a change in the plasma frequency due to the non-parabolicity of ITO conduction band. Carriers close to the bottom of the band have larger effective masses than those lying higher in the band. So in presence of a hot electrons Fermi gas there is a transient shift in the plasma frequency and, as in the inter-band case, a shift of the ITO grating photonic band gap. The result is the ultrafast photomodulation of ITO gratings spectral response in the visible by near-IR laser pulses, very interesting since they are in the telecommunication wavelengths [32].

As a second example, we discuss a totally different source of modulation of the photonic band gap: the the chamge in spectral behavior upon exposure to bacteria (*Escherichia coli*) The bacteria are acting as external stimulus in this study. The photonic structure is a simple one-dimensional photonic crystal made of silicon dioxide and titanium dioxide, covered by a thin layer of silver. The bacteria interact with silver inducing a sort of "bio-doping" of the metal. The result is a shift of the optical features of the whole silver/photonic crystal multilayer system with a consequent change of the reflected color. Such effect, that we call bacteriochromism, could be very useful for the fabrication of versatile and cheap colorimetric sensor for bacterial contamination [33,34].

## 2. METHODS

The ITO grating, with a period of 6 micrometers and duty cycle of 25% ( ITO lines width of 1.5 micrometers), has been fabricated with femtosecond laser machining. The laser system is a Yb:KGW based laser (Pharos by Light Conversion) with emitted pulses at 1030 nm, a pulse duration of 240 fs, a max. repetition rate of 1 MHz and pulse energy up to 0.2 mJ. The transmission spectra are collected with a Perkin Elmer spectrophotometer. The differential transmission pump-probe experiments have been performed with an amplified laser system based on Ti:Sapphire (Coherent Libra), with custom optical parametric amplifiers to select the excitation wavelength and an optical multichannel analyzer to collect the spectra. More details of the ITO grating fabrication and characterization are reported in Reference [32].

The $SiO_2$ / $TiO_2$ one-dimensional photonic crystals for the *Escherichia coli* optical detection have been fabricated via spin coating starting colloidal dispersion of nanoparticles. $SiO_2$ nanoparticles are purchased from Sigma Aldrich, while $TiO_2$ nanoparticles are purchased from Gentech Nanomaterials. A layer of silver (8 nm thick) have been deposited via evaporation. We have measured the transmission of the photonic crystal/silver system with a Perkin Elmer spectrophotometer after dipping them in Luria-Bertani (LB) broth only (control experiment) and in LB/*Escherichia coli* (*E. coli*) mixture. More details on the fabrication of the photonic structure and on the protocol for *E. coli* optical detection are reported in the Reference [33].

## 3. RESULTS AND DISCUSSION

In Figure 1 we show the transmission spectrum of the ITO substrate (dash-dotted curve) that we employed to fabricate the grating via femtosecond laser machining. The solid curve shows the obtained ITO grating transmission spectrum . We observe the rising of the transmission valley that can be ascribed to the photonic band gap in the periodic structure. Since the period of the grating is larger with respect to the wavelength of light, we ascribe the valley to a higher order of the photonic band gap.

In Figure 2 we present the differential transmission spectra at different time delays of the ITO grating (Figure 2A) and the differential transmission dynamics at different wavelengths of the ITO grating (Figure 2B). The excitation for this experiment is at 0.8 eV (1550 nm). We see an ultrafast modulation of the optical response of the ITO grating in the spectral region where the higher order of the photonic band gap occurs (as depicted in Figure 1).

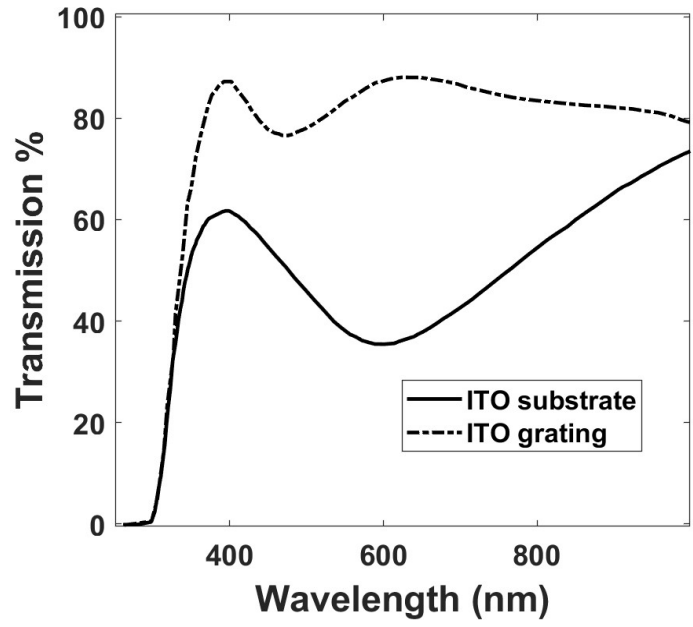

Figure 1. Transmission (%) spectra of the ITO substrate (dash-dotted curve) and of the ITO grating (solid curve).

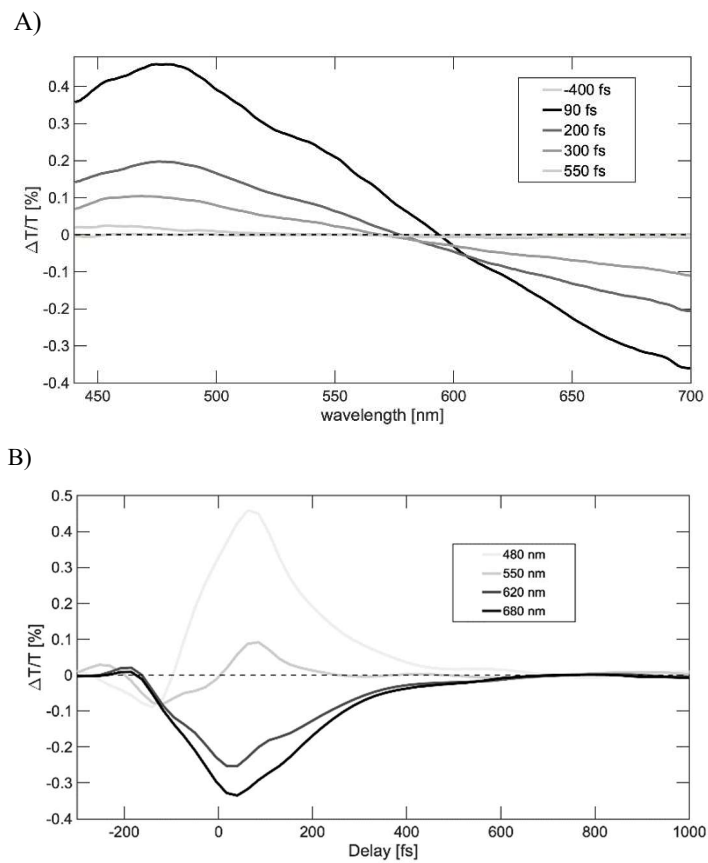

Figure 2. (A) Differential transmission spectra at different time delays of the ITO grating; (B) Differential transmission dynamics at different wavelengths of the ITO grating.

In Figure 3 we show cartoon depicting carrier distribution in ITO conduction band and the effect of non-parabolicity in the band curvature. This provides an intuitive interpretation of the reason for the shift of the photonic band gap due to near-infrared excitation. The conduction band of ITO is non-parabolic [35] and, for this reason, the hot electrons upon excitation of the Fermi gas in ITO have a lower effective mass with respect to the electrons in the Fermi gas that are close to the bottom of the conduction band. In other terms, the IR intraband excitation of the conduction band carriers allows to reduce the "average" effective mass of ITO. The different effective mass leads to a change in the plasma frequency, being the plasma frequency inversely proportional to the effective mass [25]. A change of the plasma frequency results in a change of the ITO refractive index and, thus, a change of the refractive feedback in the grating. Accordingly, we can modulate the light transmission in the visible spectral range of an ITO grating upon near-infrared excitation.

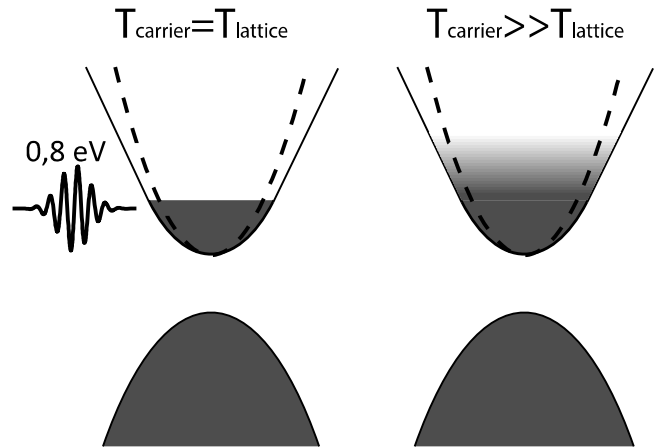

Figure 3. Sketch of the carrier distribution in the conduction band of ITO upon excitation in the near-infrared.

Above we showed that a change in the carrier effective mass, leads to a change in the grating response that is induced by a modification of the metal dielectric response. Similarly, also a change in the carrier concentration number can lead to a change in the metal dielectric response. This can be done by injecting or extracting carriers, namely by doping. Coupling with a photonic grating could amplify this effect. Here we introduce another type of hybrid plasmonic photonic structure based on $SiO_2$/$TiO_2$ photonic crystal/silver multilayer system, , This system allows detecting the presence of bacteria onto the metal surface with high sensitivity and direct color response. In Figure 4 we show the activity of the system after dipping it in LB broth (control experiment, Figure 4A) and in LB/*E. coli* mixture (Figure 4B). For an additional control experiment, we have performed the same experiment with the photonic crystal only. The spectra in Figure 4 are the average over six different samples. With respect to the red shift with LB broth only, the presence of *E. coli* determines a blue shift of the photonic band gap. This can be ascribed to a change in the carrier density of silver due to the presence of *E. coli*.

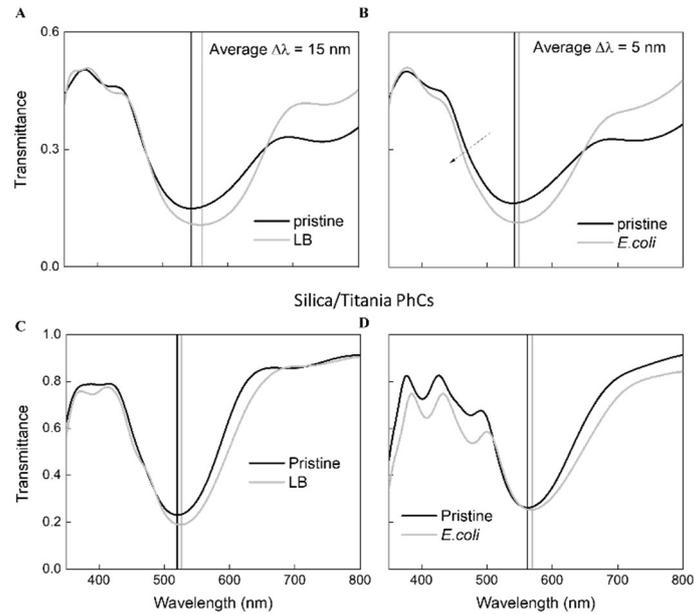

Figure 4. (A) Average transmittance spectrum of Ag/1D photonic crystals upon exposure to the LB medium and (B) to *E. coli*. (C) Average transmittance spectrum of 1D photonic crystals (without silver layer upon exposure to the LB medium and (D) to *E. coli*. Data were averaged over two sets of measurements (six samples per measurement).

## 4. CONCLUSION

In this work we show the possibility to modulate the light transmission of two different hybrid plasmonic photonic structures with external stimuli. The first system is an ITO grating that can be modulated in the ultrafast regime via excitation with near-infrared laser pulses. The second system is nanoparticle based photonic crystal covered with an 8 nm thick layer of silver that can change the position of the photonic band gap in presence of *Escherichia coli*.